\documentclass[fleqn,usenatbib]{mnras}

\usepackage{newtxtext,newtxmath}
\usepackage{hyperref}

\usepackage[T1]{fontenc}

\DeclareRobustCommand{\VAN}[3]{#2}
\let\VANthebibliography\thebibliography
\def\thebibliography{\DeclareRobustCommand{\VAN}[3]{##3}\VANthebibliography}

 \usepackage{graphicx}	
\usepackage{amsmath}	

 \usepackage{placeins}
 \usepackage{enumitem}
 
 \newlist{compactitem}{itemize}{1}
\setlist[compactitem]{
    label=\textbullet,
    topsep=0pt,
    partopsep=0pt,
    parsep=0pt,
    itemsep=0pt
}

\newcommand{\mycomment}[1]{}

\title{Analytic Modeling of CO Surface-Density Profiles in the M31 Molecular Clouds}

\author[Keto, Lada \& Forbrich]{
Eric Keto,$^{1}$\thanks{E-mail: eketo@cfa.harvard.edu (EK)}
Charles Lada,$^{2}$, Jan Forbrich$^{3}$
\\
$^{1}$Department of Astronomy, Harvard University, 60 Garden Street, Cambridge MA 02138\\
$^{2}$Center for Astrophysics, 60 Garden Street, Cambridge MA 02138\\
$^{3}$University of Hertfordshire, Center for Astrophysics Research
}

\date{Accepted XXX. Received YYY; in original form ZZZ}

\pubyear{2026}

\begin{document}
\label{firstpage}
\pagerange{\pageref{firstpage}-{0}}
\maketitle

 
\begin{abstract}
We analyze CO observations of molecular clouds in the Andromeda galaxy (M31) with a new method to derive surface-density profiles and compare directly with projected solutions of the isothermal Lane-Emden equation.  The observed curvature of the surface-density profiles is consistent with the theoretical solutions.  The applicability of the Lane-Emden equation indicates that the time scales of the turbulent processes responsible for an average force balance within the clouds are shorter than the time scales for the evolution of the clouds, for example toward collapse or disruption. The M31 profiles resemble those previously reported in Galactic molecular clouds suggesting similar dynamics across different environments. 
\end{abstract}


 \section{Introduction}\label{Intro}
 
Three recent studies introduce two new methods to analyze spectral line observations of molecular clouds to determine their internal structure. 
Applied to clouds in the Milky Way Galactic Ring (GR)
and in the Andromeda galaxy (M31),
the  new methods reveal density structure consistent with the solutions of the isothermal Lane-Emden equation 
 \citep{Keto_2024,Lada_2025,Krumholz_2025}. 
 In the GR clouds,
the internal structure  is inferred from a direct comparison, whereas in the M31 clouds,
the inference is based on the ratios of the virial energies measured as a function
of projected radius. In this study we analyze one of the new methods, differential virial analysis (DVA) introduced by \citet{Lada_2025},
and apply it to the M31 clouds to determine the radial dependence of their surface densities. 
This allows a direct comparison to the Lane-Emden solutions on par with
the previous analysis of the GR clouds and confirms the previous results for the M31 clouds. 
  
The DVA method is simple to apply, consisting of repeated
application of conventional virial analysis within progressively smaller projected areas defined by a sequence of  increasing isophote intensities.
Averages within the nested 
areas  result in data pairs of average surface density within an effective circular radius.  
The intuitive understanding of DVA is verified by application to theoretical polytropes and
model clouds from numerical simulations \citep{Krumholz_2025}.  

A statistical interpretation of the comparison between the observational results and theoretical models is complicated by correlated statistical uncertainties.  Fluctuations in the observed integrated intensity affect both the measured surface density and the position of each isophote that defines the corresponding area and effective radius. Because the isophotes are nested, these same fluctuations propagate into multiple data pairs, producing correlated uncertainties across the radial profile. In contrast, the method of \citet{Keto_2024}, which relies on azimuthal rather than areal averaging, returns the surface density directly as a function of projected radius without these covariances. Nonetheless, the covariances inherent in the DVA method are tractable, and the method is statistically well defined, as demonstrated in this study.

The benefit of the DVA method is its wider
applicability to typical radio-frequency spectral line data. The mathematically simpler method of \citet{Keto_2024} 
requires sufficient angular resolution to measure the peak integrated intensity and enough dynamic range to 
identify the half-width at half-maximum (HWHM). These conditions are not always satisfied, particularly when 
single-dish resolution is limited or interferometric dynamic range is restricted. The DVA method avoids these limitations.
One advantage of both methods is that the differential description of  the internal structure of clouds  is not 
sensitive to the absolute calibration of the CO-to-mass ratio  \citep{Krumholz_2025}.

Our analysis of the DVA method consists of a mathematical description to compare with the 
method of \citet{Keto_2024} (section \ref{CM}) and a procedure
to estimate the propagated errors that includes the covariance of surface density and radius within each data pair (appendix \ref{A}).
The procedure is simple to apply, requiring the lengths of the isophotes as 
the only additional quantity that is not already available in 
traditional observational analysis. 
We also include
a more informal analysis of the validity of the spherical approximation and the effect of the areal averaging 
(appendices \ref{space_averaging} and \ref{spherical_validity} ).  To estimate the goodness-of-fit of the
Lane-Emden model, we derive the generalized least squares $\chi^2_\mathrm{GLS}$ statistic that takes into account
the covariances of the nested data pairs.

Section \ref{THE} suggests an interpretation of the observational results: self-gravitational 
equilibrium may be established within a local region in a turbulent field if the time scale for the internal dynamics
to evolve to a balance of forces is shorter than the time scale for disruption or collapse of
the cloud by self-gravity or the larger-scale surrounding turbulence.

Comparison of the analyses of the extragalactic clouds in M31 with the previously analyzed 
GR clouds (section \ref{GEG}) suggests similar dynamical states in both populations.

\section{Description of the method}\label{CM}

\citet{Keto_2024} defines molecular clouds as regions of higher surface density around a local emission peak.
The
azimuthally-averaged radial profile of the surface density, $\Sigma(r)_{\rm R}$ is,
\begin{equation}\label{obsR}
\Sigma(r)_{\rm R} = \frac{1}{2  \pi r \Delta R} \int_0^{2\pi} \int_{r-\Delta R/2} ^{r+\Delta R/2} 
N(r^\prime\cos\theta,r^\prime\sin\theta) r^\prime dr^\prime d\theta,
\end{equation}
where $N$ is the observed surface density defined as a function of the sky coordinates $(x,y)$, 
and $r$ is the projected
radius from the peak surface density that is defined as the center of the cloud. 
The subscript R on the observed surface density $\Sigma$ indicates that this definition is based on the projected radius.

Assuming that the $^{12}$CO(2-1) integrated intensity is proportional to the mass surface density (\S \ref{Xfactor} ),
the corresponding theoretical surface density, $S$,  for comparison with the solution of the Lane-Emden equation
 is,
\begin{equation}\label{LER}
S_R(x) = 2 \int_0^{\sqrt{R_0^2 - x^2}} n(\sqrt{x^2 + z^2})\ dz,
\end{equation}
where  $R_0$ is the truncation radius, $n$ is the number density,  
$y=0$, and $z$ is coordinate along the line of sight.
The observations of the GR clouds indicate that in practice, $R_0$ is large
compared to the characteristic size of the clouds as measured by their half-width
at half-maxima (HWHM).
The observational and theoretical profiles of surface density may be compared in
non-dimensional units by scaling an
observed profile by its peak density and its width by its HWHM.

\citet{Lada_2025} define a molecular cloud as a region of higher surface density
enclosed within an isophote. 
For each isophote or  threshold $N_i$, the set of all pixels satisfying
\begin{equation}\label{Set}
\Omega(N_i) = \{ (x,y) | N(x,y) > N_i \}
\end{equation}
defines the region $\Omega(N_i)$. 
The average surface density within the isophote $N_i$ is,
\begin{equation}\label{ASD}
\Sigma_A(N_i) = \frac{1}{A(N_i)} \int_{\Omega(N_i)} N(x,y) dxdy,
\end{equation}
where the area, $A(N_i)$, is the number of pixels in the set $\Omega(N_i)$ times the pixel area $dxdy$.
Each area-based average surface density $\Sigma_A(N_i)$ is
associated with a corresponding effective radius $r_i = \sqrt{ A(N_i)/\pi}$. 
We can refer to $S_A(N_i)$ as $S_A(r_i)$. 
The subscript $A$ on the surface density
indicates that this definition is based on the area within an isophote.

The corresponding surface density from the Lane-Emden equation is defined as,
\begin{equation}\label{LEA}
S_{\rm A}(r_i) = \frac{1}{\pi r_i^2} \int_0^{r_i} S_R(r^\prime)\ 2\pi r^\prime dr^\prime .
\end{equation}

In the DVA method, the scaling procedure to compare the area-based observational and theoretical profiles is necessarily
different than the scaling procedure used in \citet{Keto_2024} because the cloud center, the peak surface density,
and the HWHM are not observationally defined.
For each cloud, the set of observed surface densities $\Sigma_{\rm A}(r_i)$  represents
a segment of the complete radial profile. The segment can be 
matched to the theoretical profile $S_{\rm A}(r)$ as a function of
the position.  At any trial position, 
we have a single vertical and a single horizontal scaling
factor to convert the non-dimensional theoretical profile to observational units. 
This procedure fits the {\it curvature} of the observed and theoretical profiles. Since the profile of the
surface density derived from the Lane-Emden solution can be approximated as two segments with different simple power laws, the fit
is unique only if it includes the transition region between these power-law regions where the curvature is changing
rather than constant. In contrast, the method of \citet{Keto_2024} avoids this limitation. It simply compares the normalized profiles
without the requirement of fitting.

\section{Results}\label{Results}

FIgure 1 compares projected solutions of the isothermal Lane-Emden equation for hydrostatic equilibrium to the observed radial surface density profiles of 24 of the 26 M31 clouds from C.J. Lada et al (2025) that contain at least 10 radial points. 
Two clouds (K297A and K301A) lacked sufficient curvature to obtain a unique fit because their observed points lie entirely in the inner or outer power-law regions of the profiles and are not shown here. For each of the remaining 24 clouds the figure shows only every other point for clarity. The error bars on the plots are calculated according to the procedure described in appendix \ref{A}. Also shown on the plots are
the reduced $\chi^2_\mathrm{GLS}$ values for the model fits.  

We evaluate the fits using the generalized least-squares statistic, $\chi^2_\mathrm{GLS}$, which accounts for the full covariance matrix of the surface-density measurements as described in Appendix~\ref{gls}. For 23 out of 24 clouds, the reduced statistic $\chi^2_\mathrm{GLS}/\nu$, where $\nu = N - k$ is the number of degrees of freedom, is of order unity. This indicates that the residuals are consistent in magnitude with the observational uncertainties and that no systematic deviations from the Lane-Emden model are detected. The $\chi^2_\mathrm{GLS}/\nu$ values therefore provide a quantitative confirmation of the close agreement between the observed profiles and the predictions of the Lane-Emden equation for the clouds as shown in the figure.

\citet{Keto_2024} previously found that the 
surface density profiles of the GR clouds match the projected solutions of the 
Lane-Emden equation at all observed projected distances. 
The same result obtained here for a different set of clouds and by a
different 
analytic method confirms that the results are not dependent 
on the analytic method. The comparison indicates that the two populations share similar dynamics.

\citet{Lada_2025} previously found that virial equilibrium applies throughout the M31 clouds when the
projected virial volume is a function of projected radius following the DVA method.
At larger radii, the balance
of energies is dominated by the kinetic and gravitational energies. At smaller radii where the
gravitational energy decreases along with the enclosed mass, the balance shifts to a dominance
by the kinetic energy inside the virial volume and the pressure energy at its surface. 
The balance of forces defined by the Lane-Emden equation is consistent with this shift in 
dominant energies.

\section{Interpretation}

If the M31 clouds are described by the Lane-Emden equation, then we can infer certain conditions on the dynamical state of the turbulent clouds from the 
physical state described by the equation.
The Lane-Emden equation combines the equation for hydrostatic equilibrium with a polytropic equation of state. 
The hydrostatic condition limits the significance of any acceleration of the gas within the cloud with respect to the
inward force from self-gravity and the outward force of the pressure gradient. The limit can be derived from the 
Euler equation for an inviscid fluid with self-gravity,
\begin{equation}
\rho \frac { D {\bf v}  }   { Dt  } = -{\bf \nabla}P - \rho {\bf \nabla}\Phi
\end{equation}
where $D/Dt  =\partial_t + {\bf v \cdot \nabla}$ is the material derivative and the terms
on the right hand side are the forces due to pressure and gravity. 
For a characteristic length scale $L$ and fluid speed $V$, the magnitude of the acceleration term is,
\begin{equation}
\left |  \rho \frac { D{\bf v  } }   { D t  }  \right |
\sim \rho \left ( \frac {V}{t} + \frac{V^2}{L}  \right ) 
\sim \rho \frac {V^2} {L} ,
\end{equation}
where the second equality follows if the flow varies on a characteristic timescale $\sim L/V$.
We can estimate the magnitudes of the pressure and gravitational forces as,
\begin{equation}
| {\bf \nabla} P | \sim \frac {\Delta P} {L} \quad \mathrm{and} \quad \rho | {\bf \nabla} \Phi | \sim \rho g .
\end{equation}
This limits the large-scale or bulk velocities in the clouds, 
\begin{equation}
V^2 \ll \frac {\Delta P} {\rho} \quad \mathrm{and} \quad V^2 \ll g L .
\end{equation}
The comparison in figure 1 also indicates that the equation of state is effectively isothermal so that
$P \approx \rho a^2$ where $a$ is a constant effective sound speed. Therefore, 
$V \ll a$. This limits the bulk flow within the clouds to speeds less than the 
effective sound speed which is
approximately the turbulent velocity dispersion. 
The second condition also limits the bulk flow speed to less than the  
characteristic dynamical speed 
$v_\mathrm{dyn} \equiv \sqrt{g L}$, for example the free-fall speed or the virial speed.

We can also derive equivalent conditions on the time scales. Define the sound crossing and
free-fall time scales as,
\begin{equation}
t_x \sim \frac {L}{a} \quad \mathrm{and} \quad t_\mathrm{dyn} \sim \sqrt{ \frac{L}{g}  }.
\end{equation}
Then $V \ll a$ and $v \ll \sqrt{gL}$ require that 
$t_\mathrm{evo} \gg \mathrm{max} [t_x, t_\mathrm{dyn}] $. Therefore, figure 1 
indicates that the 
evolutionary time scales of the large-scale flows, $t_\mathrm{evo}$, that lead to  the gravitational collapse or disruption 
of the M31 clouds are long compared the time scales to establish a balance between 
pressure and self-gravity.

The conditions required by the Lane-Emden equation are independently supported by the observations.
The observed line widths are approximately
constant within each cloud indicating that the turbulent kinetic energy is effectively isothermal. 
The observed mean velocities of the 
CO spectral lines vary with position in the clouds by less than the observed line widths.
This implies that at least along the line-of-sight, the bulk flow speeds are less than
the line widths. Spectral line profiles such as asymmetric split
lines indicative of collapse or expansion velocities greater than the line width are not observed.

\section{Discussion} \label{THE}

Consider a cloud  defined as in section \ref{CM} by the extent of a high-density region in a field of self-gravitating turbulence.
The self-gravitating cloud is continuously disturbed by turbulence on all scales, interior and exterior. 
The mean dynamical state of the cloud is defined by a 
spatial average over the fluctuating turbulent substructure on scales up to its characteristic dynamical length,
for example a Jeans length. The observed state of the cloud represents a mean state after the averaging that is 
inherent in the observation
and applied in the 
data analysis smooths out the local fluctuations. 

The observational comparison with the Lane-Emden solutions requires the following conditions apply to the M31 clouds. \\
1) The mean state is one of approximate self-gravitational equilibrium. \\
2) The turbulent processes responsible for establishing the force balance between the pressure gradient and self-gravity
operate on a shorter time scale than the turbulent processes responsible for the collapse or disruption of the cloud. 

Following \citet{Keto_2024}, we can understand this separation of time scales from the spatial dependence of the turbulent dynamical time scales.
The turbulent processes within a cloud that are responsible for its internal structure, operate on a range of time scales
with a {\it maximum} defined by the crossing time of the cloud. The turbulent fluctuations that
are of large enough scale to modify the entire cloud operate on a range of time scales with a {\it minimum} defined
by the same crossing time of the cloud. This separation of time scales allows a cloud  
to evolve toward self-gravitational equilibrium within the stochastic but relatively slowly changing boundary conditions of
the local environment defined by the surrounding, larger-scale turbulence.

Apparent equilibrium is found to describe many observations
of molecular clouds. For example, Galactic and extragalactic 
surveys covering galactic scales  
 find average virial parameters
in the range of (1 - 2) \citep{Heyer_2009,Roman-Duval_2010,Miville_2017,Evans_2021,Wong_2011,Colombo_2014,Hughes_2013}. 
Occasional values $\sim 3$ are explainable within a 
higher pressure environment.
These observational surveys reveal mean dynamical states with a high degree of correlation with states of hydrostatic or virial equilibrium
over a wide range of cloud scales. This suggests that the equilibrium mean state is an endpoint of a deterministic evolutionary process operating
within self-gravitating turbulence.

\subsection{Comparison of galactic and extragalactic clouds}\label{GEG}

The earlier analysis of the GR clouds  resulted in four main observational findings \citep{Keto_2024}. 
{\setlength{\parskip}{0pt}
\begin{enumerate} [nosep]
 \item
The surface densities  are consistent with solutions of the isothermal Lane-Emden equation.
\item
The clouds are in approximate virial equilibrium with an external pressure that is different around each cloud. 
\item
The required external pressure matches the turbulent pressure of the surrounding molecular ISM. (In the M31 clouds this pressure
is equal to the external pressure in the virial theorem which indicates equilibrium at all radii.)
\item
The mid-plane pressure of the multi-phase turbulent ISM acts as an external boundary
condition for the larger-scale, turbulent molecular ISM.
\end{enumerate}
From the earlier virial analysis in \citet{Lada_2025} and the analysis with the Lane-Emden equation presented here,
these four descriptions of the GR clouds apply to the extragalactic M31 clouds as well. The two populations therefore share
similar dynamics.
} 

\section{Conclusions}
We have analyzed the CO emission from molecular clouds in M31 using the area-based DVA method of Lada et al. (2025) in which effective radii and corresponding surface densities are derived from nested isophotal regions. When the same averaging procedure is applied to the isothermal Lane-Emden solutions, the theoretical and observed profiles can be compared directly. For 24 of 26  clouds with sufficient radial sampling, the observed surface-density profiles agree closely with the Lane-Emden predictions. The observations imply that self-gravitational equilbrium can develop within
a turbulent field because the time scale for the establishment of force balance on the spatial scales within the clouds is shorter than the time scale for their collapse or disruption by larger scale turbulent fluctuations on spatial scales outside the clouds.

The close similarity between the M31 clouds studied here and previously analyzed in the Galactic ring shows that both populations share comparable dynamical states.

We provide a statistical interpretation that accounts for the inherent covariances in the DVA method.
We define a
simple procedure for the approximate propagation of errors, and we derive the generalized least squares $\chi^2_\mathrm{GLS}$ statistic
for the estimation of the goodness-of-fit with theoretical models. We analyze the spherical approximation and find justification if the effective support (thermal plus turbulent plus magnetic) acts isotropically on average, and the gravitational potential is dominated by a centrally condensed mass even if individual isophotes deviate from circularity. We consider how area-based averaging further suppresses local fluctuations, yielding smooth profiles that represent a mean structure on the local scale.

\section{Data Availability}
This study uses data published with \citet{Lada_2025}. No new data were 
generated in this study.

\bibliography{GRS_2024_08_27_REV}{}
\bibliographystyle{aasjournalv7}

\appendix
\section{Estimated uncertainties}\label{A}

Here we
express the variances and covariance of the surface density and radius in terms of those measured
for the mass and area. We follow the method outlined in Section~3.2 of \citet{Lada_2025} to derive the variances of
the area and mass within each isophote (equations \ref{A1} and \ref{A2}). Because the location of an isophote depends on random
fluctuations in the observational noise, we also derive the corresponding covariance. 

The mass is defined as the area integral of the observed surface density $N$ over the isophotal
region defined by equation~\ref{Set},
\begin{equation}
M_i = \int_{\Omega(N_i)} N\, dA .
\end{equation}
From equation \ref{Set}, the index $i$ labels the nested cumulative isophotal regions. 
The average gradients of area and mass around the isophote are
\begin{equation}\label{A1}
A'_{\rm i} = \frac{dA}{dN}\bigg|_{N_{\rm i}}
\approx \frac{A_{\rm i+1} - A_{\rm i-1}}{\Delta N},
\end{equation}
\begin{equation}\label{A2}
M'_{\rm i} = \frac{dM}{dN}\bigg|_{N_{\rm i}}
\approx \frac{M_{\rm i+1} - M_{\rm i-1}}{\Delta N},
\end{equation}
where $\Delta N = N_{\rm i+1} - N_{\rm i-1}$.
These gradients represent the change in the values of the area and mass
due to a shift of the isophote caused by observational
noise. Because these estimates are based on the actual shape of the isophote,
they empirically include the added uncertainty due to non-circular
boundaries.
Because the gradients represent a coherent shift
whereas random fluctuations along the isophote partially cancel, the variances derived from
these estimates include a factor for the correlation length,
\begin{equation}\label{A3}
\lambda_i =  \sqrt{\frac{\ell_c}{L_{\rm i}}},
\end{equation}
where  the correlation length is
$\ell_c = \sqrt{2}\,\mathrm{FWHM}$, and $L_{\rm i}$ is the isophote length.
The square root factor follows from the random-walk nature of the fluctuations.
The variances are
\begin{equation}\label{A4}
{\rm Var}(A_{\rm i}) = (A'_{\rm i})^2 \lambda_i^2
\sigma_{\rm b}^2,
\end{equation}
\begin{equation}\label{A5}
{\rm Var}(M_{\rm i}) =
A_{\rm i} \sigma_{\rm b}^2 \Omega_{\rm b}
+ (M'_{\rm i})^2 \lambda_i^2
\sigma_{\rm b}^2,
\end{equation}
\begin{equation}\label{A6}
{\rm Cov}(M_{\rm i},A_{\rm i}) =
A'_{\rm i} M'_{\rm i}\, \lambda_i^2
\sigma_{\rm b}^2,
\end{equation}
where $\Omega_{\rm b}$ is the beam area, $\sigma_{\rm b}$ is the observational noise, and the first term in ${\rm Var}(M_{\rm i})$
is the contribution from observational noise within the isophote.

We map the measured uncertainties in area and mass to uncertainties in surface density,
$\Sigma = M/A$, and radius, $r = \sqrt{A/\pi}$. 
The surface density and radius are correlated variables, and the statistical uncertainty in
the surface density includes terms arising from the covariance with radius.
For the vector of random variables
${\bf X} = (A,M)$ and the vector function ${\bf Y} = (\Sigma,r)$, the covariance matrix
${\bf C}({\bf Y})$ is obtained from ${\bf C}({\bf X})$ through
\begin{equation}\label{A7}
{\bf C}({\bf Y}) = {\bf J}\,{\bf C}({\bf X})\,{\bf J}^{\rm T},
\end{equation}
where the Jacobian is
\begin{equation}\label{A8}
{\bf J} =
\left[
\frac{\partial\Sigma}{\partial A} = -\frac{\Sigma}{A},\quad
\frac{\partial\Sigma}{\partial M} = \frac{1}{A},\quad
\frac{\partial r}{\partial A} = \frac{1}{2\sqrt{\pi}\,r},\quad
\frac{\partial r}{\partial M} = 0
\right].
\end{equation}
The variances for $\Sigma$ and $r$ are then
\begin{equation}\label{A10}
{\rm Var}(\Sigma) =
\frac{1}{A^2}{\rm Var}(M)
+ \frac{\Sigma^2}{A^2}{\rm Var}(A)
- \frac{2\Sigma}{A^2}{\rm Cov}(M,A),
\end{equation}
\begin{equation}\label{A10}
{\rm Var}(r) = \frac{1}{4\pi A}\,{\rm Var}(A).
\end{equation}
The covariance and correlation of $\Sigma$ and $r$ are
\begin{equation}\label{A11}
{\rm Cov}(\Sigma,r) =
\frac{1}{2\sqrt{\pi A}}\,
\left[{\rm Cov}(M,A) - \Sigma\,{\rm Var}(A)\right],
\end{equation}
\begin{equation}\label{A12}
\rho = \frac{{\rm Cov}(\Sigma,r)}{\sigma_\Sigma \sigma_r}.
\end{equation}

The uncertainty of $r_i$ (equation~\ref{A10}) depends on the gradient of the area
(equation~\ref{A1}), rather than the localization error associated with the finite beam width.
Because 
the gradient is measured from
the beam-smoothed image, the observed variance of the area includes the
effect of the finite resolution.

As is often the case in radio-frequency
observations, particularly interferometric observations, the peak
signal-to-noise ratio exceeds the spatial dynamic range. Consequently,
isophotes separated in value by the observational noise typically lie
closer together than the beam width ($2.3 \times 1.9^{\prime\prime}$
or $8.5\times 7$  pc HWHM). In the M31 sample, the estimated error in the
area-based radius is smaller by a factor of order ten  
than the localization uncertainty set by the finite resolution.
Intuitively, the error in the area-based radius reflects the average over
many independent beams within an isophote, whereas the resolution-based
uncertainty derives from a single measurement. The uncertainties from the
covariance depend on the variance of the area or radius and are small
compared to the variance in the surface density due to observational
noise interior to the isophote.

\section{Smooth, spherical clouds}

\mycomment{
The smoothness of the radial surface density profiles in figure 1 may be
surprising given the irregular images of most molecular clouds. The smoothness
is the result of averaging  inherent in the DVA analysis. The two averages that
are the most 
significant, the averaging of non-circular geometries to circular and the averaging
of the proportionality of the CO integrated intensity to mass surface density are
discussed below.

The applicability of the spherical Lane-Emden equation to irregular molecular
clouds depends on two properties that are related to the sphericity of clouds.
These are the directions of the vector of the gravitational force and the vector
of the pressure gradient. The required conditions are discussed in the third section
below.
}

\subsection{Averaging of non-circular geometries to circular}\label{space_averaging}

The CO integrated intensities of clouds in the images of \citep{Lada_2024} show
fluctuations of the isodensity contours  (isophotes) that are not evident in the 
smoothness of the observed
surface density profiles. The smoothness in the profiles is a result of
three averages that apply in the analytic procedure described in section \ref{CM}.

(1) The surface density at an effective radius is the average over all pixels within the corresponding isophote 
rather than the average of values at a single radius.

(2) The effective radius depends only on the area
and not on the perimeter geometry.

(3) Successive values of the surface density and effective radius
are not independent since each value incorporates the areas and surface densities of all nested higher-threshold isophotes.

\subsection{Validity of spherical approximation for non-circular clouds}\label{spherical_validity}

The Lane-Emden equation is spherically symmetric while
some of the isophotes in the images of the M31 molecular clouds are
non-circular \citep{Lada_2024}. 
The validity of the spherical approximation requires two conditions. 
 
(1) The effective support (thermal + turbulent + magnetic) must be approximately isotropic or radial.

(2) The gravitational potential must be dominated by the cloud's own mass distribution, and  the distribution must be centrally condensed. 

These two conditions generally hold when the spectral lines show a constant or radially dependent velocity dispersion,
and the surface density has a single maximum, explained further below.

(1) With the internal energy dominated by turbulence and the approximately constant velocity dispersion of the CO lines within each cloud,
the effective pressure  depends on the local scalar $\rho a^2$,
where $a$ is an effective sound speed that includes non-thermal contributions.  
The pressure gradient then follows
the density gradient. In the images, the isophotes mark thresholds in the continuous field
of the surface density rather than discontinuities. Therefore, the pressure varies
smoothly across the isophotes and on average is dominated by its radial component
even if an isophote is not everywhere perpendicular to the radial direction.

(2) The gravitational potential is derived from the Poisson integral of the density.
The integration effectively smooths over perturbations of the density structure.
Higher-order multipoles from non-spherical shapes, evidenced by elongated or locally fluctuating isophotes,
contribute at order-unity
while the monopole from the enclosed mass remains dominant.

\subsection{Proportionality of the CO luminosity and the mass surface density of clouds}\label{Xfactor}

Observations of $^{12}$CO have provided most of the estimates of the surface density
and mass of the molecular ISM over the past 50 years.
The empirically based proportionality of the $^{12}$CO luminosity or integrated intensity and the mass or
mass surface density has been extensively studied and verified as generally correct over a wide range of cloud 
masses and column densities \citep{Bolatto_2013}. 
We use the conversion factor of $\alpha_{\rm ^{12}CO(2-1)} = 10 \pm 4.5 $ M$_\odot$ (K km s$^{-1}$ pc$^2$)$^{-1}$
determined specifically for $^{12}$CO(2-1) in M31 by direct comparison with dust emission \citep{Lada_2024}.
As noted in the introduction, our results depend only on the proportionality rather than the value of the scaling factor.

\section{Generalized Least-Squares Fit to the Surface Density Profile}
\label{gls}

The surface-density profile derived from nested isophotes
produces a sequence of values
\begin{equation}
\boldsymbol{\Sigma}
=
(\Sigma_1,\Sigma_2,\dots,\Sigma_N)^T,
\end{equation}
where
\begin{equation}
\Sigma_i = \frac{M_i}{A_i},
\qquad
r_i = \sqrt{\frac{A_i}{\pi}}.
\end{equation}
Because each isophote encloses all higher-threshold regions,
successive values are not statistically independent.
Therefore, the appropriate goodness-of-fit statistic
for comparison to a model profile
$\Sigma_{\rm model}(r)$
is the generalized least-squares (GLS) chi-square,
\begin{equation}
\chi^2_{\rm GLS}
=
\boldsymbol{\Delta}^T
\mathbf{C}_\Sigma^{-1}
\boldsymbol{\Delta}
\label{eq:chi2_gls}
\end{equation}
The elements of the residual vector $\bf \Delta$ are
\begin{equation}
\Delta_i
=
\Sigma_i
-
\Sigma_{\rm model}(r_i) .
\end{equation}
The full covariance matrix of the
surface-density vector $\boldsymbol{\Sigma}$ is, 
\begin{equation}
(\mathbf{C}_\Sigma)_{ij}
=
\mathrm{Cov}(\Sigma_i,\Sigma_j).
\end{equation}

Using standard first-order error propagation,
for any two contour levels $i$ and $j$,
\begin{equation}
\mathrm{Cov}(\Sigma_i,\Sigma_j)
=
\sum_{\alpha,\beta}
\frac{\partial \Sigma_i}{\partial X_{i,\alpha}}
\frac{\partial \Sigma_j}{\partial X_{j,\beta}}
\,
\mathrm{Cov}(X_{i,\alpha},X_{j,\beta}),
\end{equation}
where
\begin{equation}
X_{i,\alpha} \in \{M_i,A_i\}.
\end{equation}

The required derivatives are

\begin{equation}
\frac{\partial \Sigma_i}{\partial M_i}
=
\frac{1}{A_i},
\qquad
\frac{\partial \Sigma_i}{\partial A_i}
=
-\frac{M_i}{A_i^2}
=
-\frac{\Sigma_i}{A_i}.
\end{equation}

Thus,

\begin{align}
\mathrm{Cov}(\Sigma_i,\Sigma_j)
&=
\frac{1}{A_i A_j}
\mathrm{Cov}(M_i,M_j)
\nonumber\\
&\quad
-
\frac{\Sigma_j}{A_i A_j}
\mathrm{Cov}(M_i,A_j)
\nonumber\\
&\quad
-
\frac{\Sigma_i}{A_i A_j}
\mathrm{Cov}(A_i,M_j)
\nonumber\\
&\quad
+
\frac{\Sigma_i \Sigma_j}{A_i A_j}
\mathrm{Cov}(A_i,A_j).
\label{eq:cov_sigma_full}
\end{align}

\subsection{Covariance of Nested Isophote Quantities}

Because the isophotes are nested,
the enclosed quantities satisfy

\begin{equation}
A_i = \sum_{k =0}^{i} \delta A_k,
\qquad
M_i = \sum_{k =0}^{i} \delta M_k,
\end{equation}

where $\delta A_k$ and $\delta M_k$
represent the area and mass in the disjoint annular region
between successive contour levels.
Assuming independent noise in disjoint annuli,
the covariances of the cumulative quantities are

\begin{align}
\mathrm{Cov}(A_i,A_j)
&=
\sum_{k=0}^{\min(i,j)}
\mathrm{Var}(\delta A_k),
\\
\mathrm{Cov}(M_i,M_j)
&=
\sum_{k=0}^{\min(i,j)}
\mathrm{Var}(\delta M_k),
\\
\mathrm{Cov}(M_i,A_j)
&=
\sum_{k=0}^{\min(i,j)}
\mathrm{Cov}(\delta M_k,\delta A_k).
\end{align}

These expressions explicitly encode the correlation introduced
by nested isophotes.
Substituting these into Equation~(\ref{eq:cov_sigma_full})
yields the full covariance matrix $\mathbf{C}_\Sigma$.

\subsection{Statistical Interpretation}

The analytic methods of error estimation in appendices A and C are approximate methods
developed specifically for the DVA technique. The most reliable method of error estimation
is a Monte Carlo simulation embedded in the data analysis pipeline.

Appendix A derives the estimated uncertainties that apply to 
each individual $(r,\Sigma)$ pair. Appendix C estimates the goodness-of-fit
that applies to the comparison of the model with all the data points for each cloud
including the effect of their covariances.

The weighted least squares statistic, $\chi^2_{\rm diag}$, from the diagonal terms of
the covariance matrix, can be predicted from figure 1 according to how well
the model lies within the error bars of the data. Although, the $\chi^2_{\rm GLS}$ similarly assesses
the likelihood of the data given the model, the effect of the covariances
is not obvious from the figure.

For example, 
consider data that is consistently above or below the model curve.
If the covariances are strong implying that
the data points are tied together rather than independent, then the
likelihood of the data given the model may be higher than would
be the case if the data points were more independent, with weaker covariances. This is because
the mismatch may be due to a single observational
error that affects all the data points.  Conversely, in the case of data that provide a good
fit on average but wander above
and below the model curve, strong covariances might imply a lower likelihood
if the
observational errors must be arranged in the particular pattern of the data.

In the ideal limit,  the $\chi^2_{\rm GLS}$ follows
a $\chi^2$ distribution with $\nu$ degrees of freedom where $\nu = n - k$
for a model with $n$ data points and $k$ free parameters.
However, this assumes the following conditions: 
1) the residual vector must be approximately multivariate Gaussian;
2) the covariance matrix $C_\Sigma$ must be known, rather than estimated from the data;
3) the fitted model must be sufficiently regular, so that subtracting $k$ fitted parameters is the right degrees-of-freedom correction;
4) the clouds pooled into a sample must all share the same $\nu$ and the same covariance quality.

Given these considerations, the statistics support the conclusion that the
Lane-Emden model provides a satisfactory description of the observed profiles.
For most  clouds, 22 out of 24, $\chi^2_{\rm GLS} > \chi^2_{\rm diag}$.
This may reflect subtle curvature mismatches between model and data, or limitations in the approximate covariance model.
Nevertheless, $\left \langle \sqrt{ { \chi^2_{\rm GLS} }  / {\nu } } \right \rangle  = 0.44  \pm 0.20 $,
and $\left \langle \sqrt{ { \chi^2_{\rm diag} }  / {\nu } } \right \rangle  = 0.37  \pm 0.22 $, both
below the ideal value of unity for an exactly calibrated uncertainty model.
This indicates that the fits are satisfactory but that the adopted uncertainties are conservative, with the dominant overestimate likely originating in the uncertainties assigned to the individual
$(r,\Sigma)$ data pairs and then propagating into the estimated covariances. 

\begin{figure*}
\begin{tabular} {lcr}
\includegraphics[width=2.5in,trim={0 0.0in 0.0in 0.0in},clip]{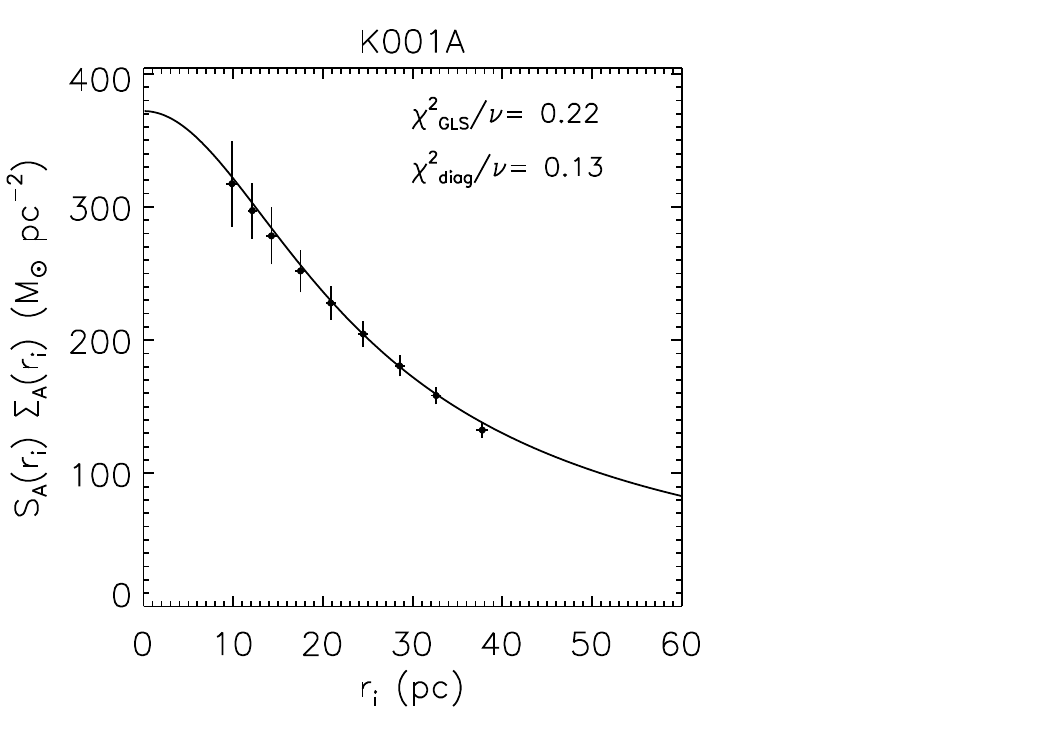} &
\includegraphics[width=2.5in,trim={0 0.0in 0.0in 0.0in},clip]{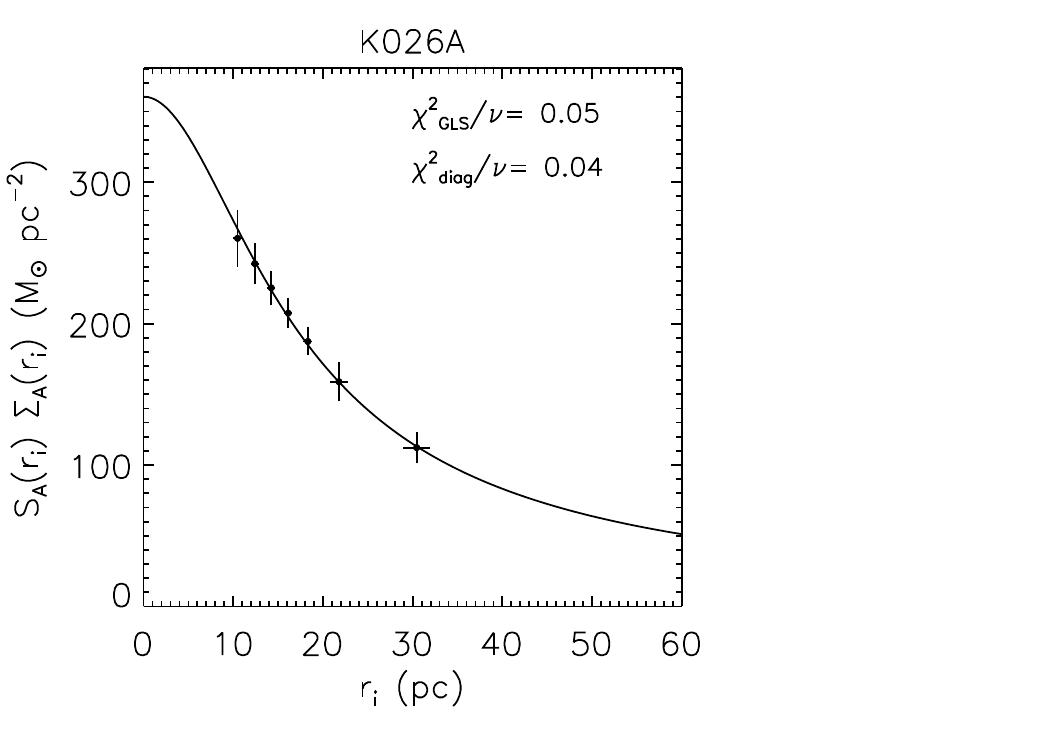} &
\includegraphics[width=2.5in,trim={0 0.0in 0.0in 0.0in},clip]{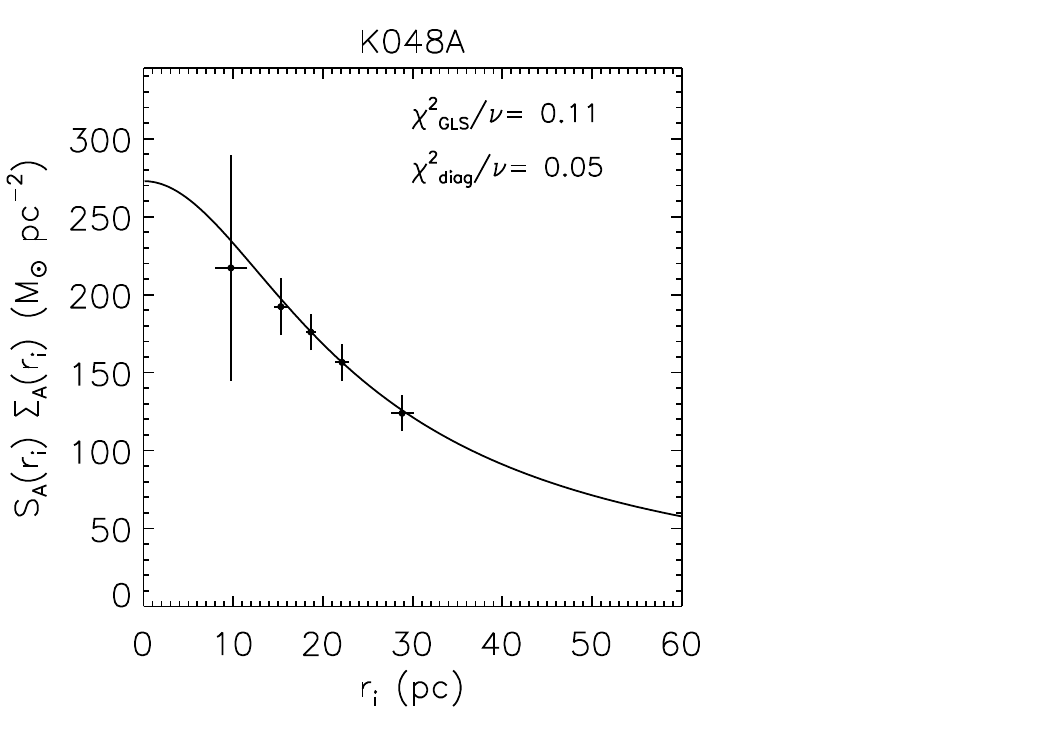} \\
\includegraphics[width=2.5in,trim={0 0.0in 0.0in 0.0in},clip]{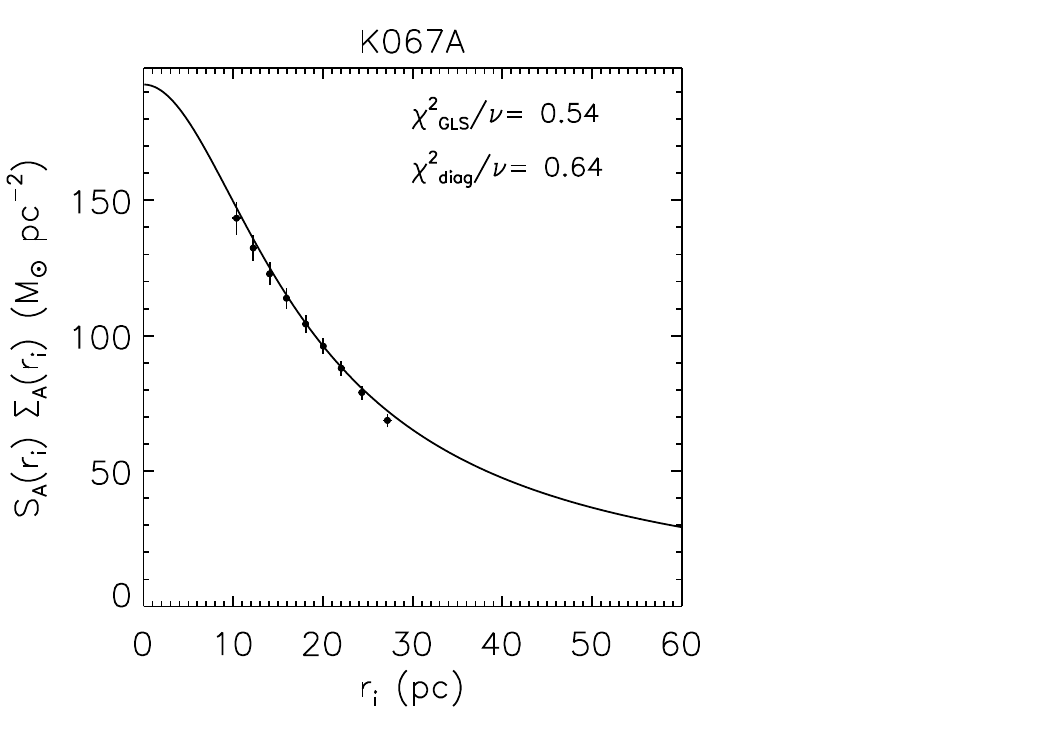} &
\includegraphics[width=2.5in,trim={0 0.0in 0.0in 0.0in},clip]{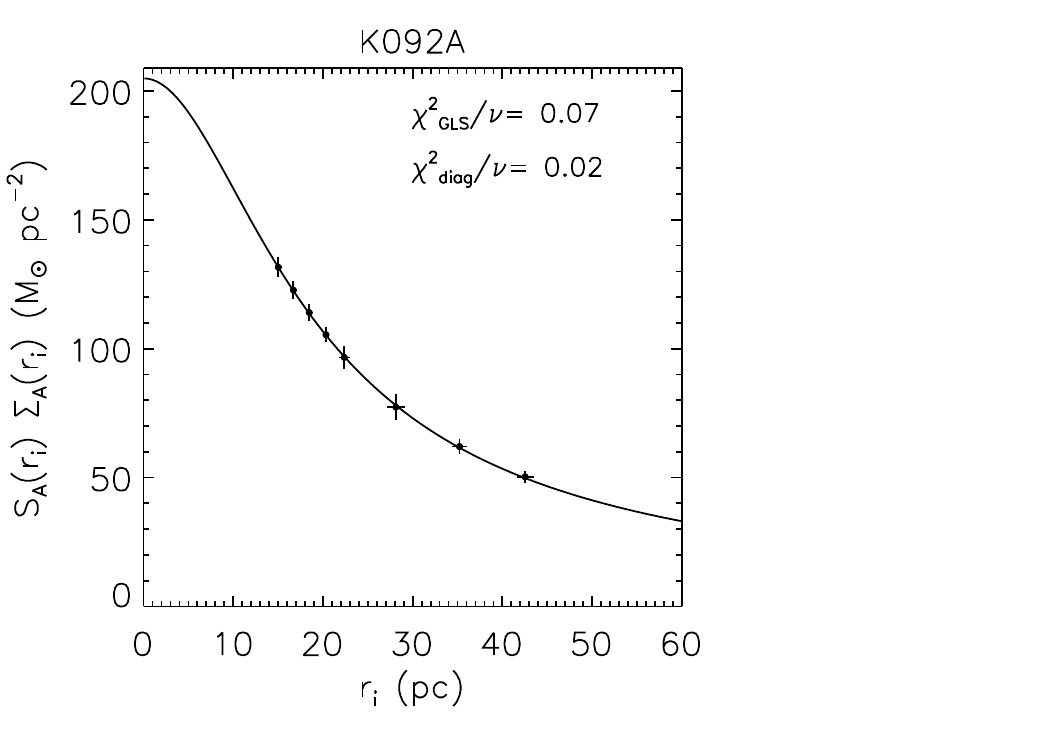} &
\includegraphics[width=2.5in,trim={0 0.0in 0.0in 0.0in},clip]{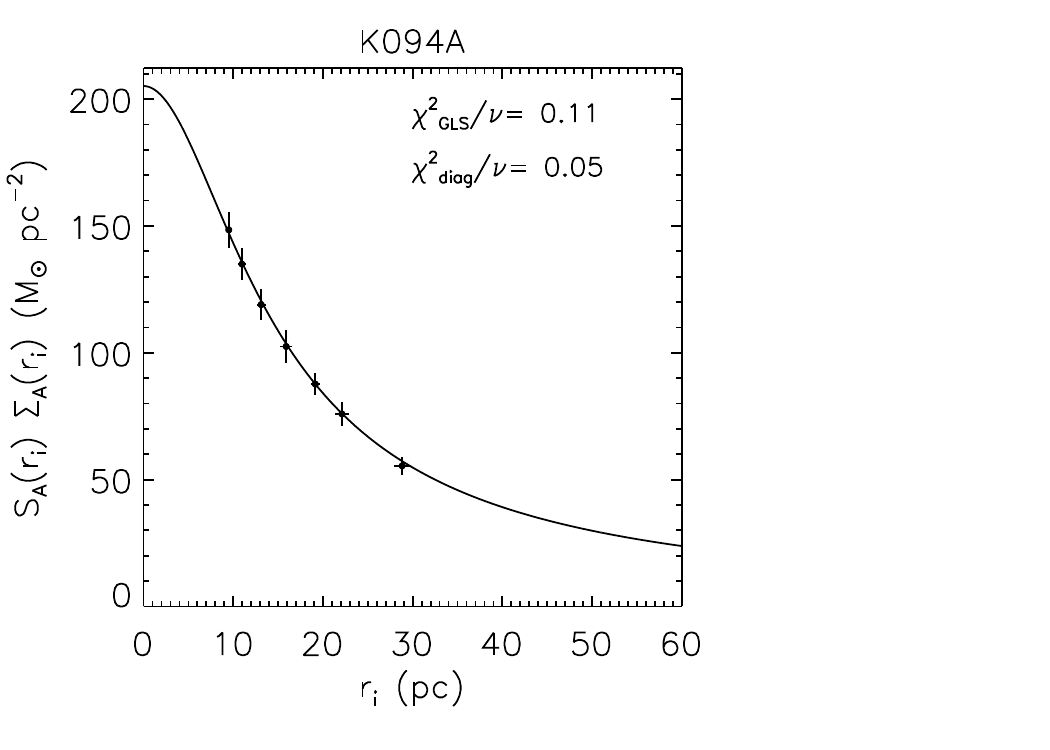} \\
\includegraphics[width=2.5in,trim={0 0.0in 0.0in 0.0in},clip]{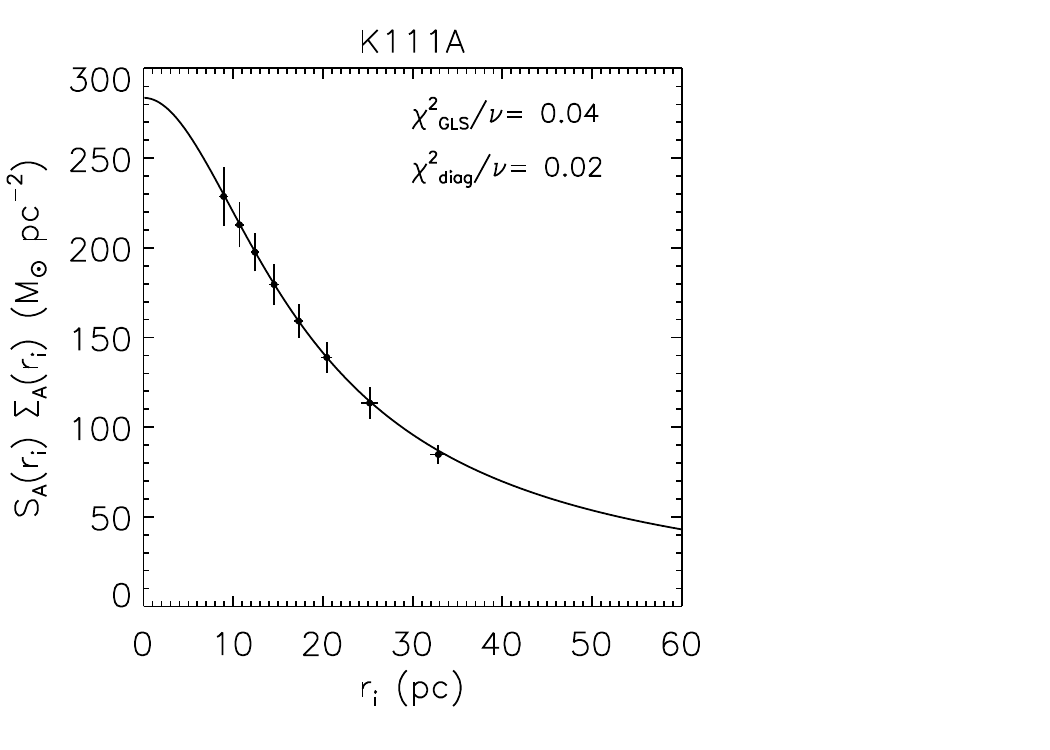} &
\includegraphics[width=2.5in,trim={0 0.0in 0.0in 0.0in},clip]{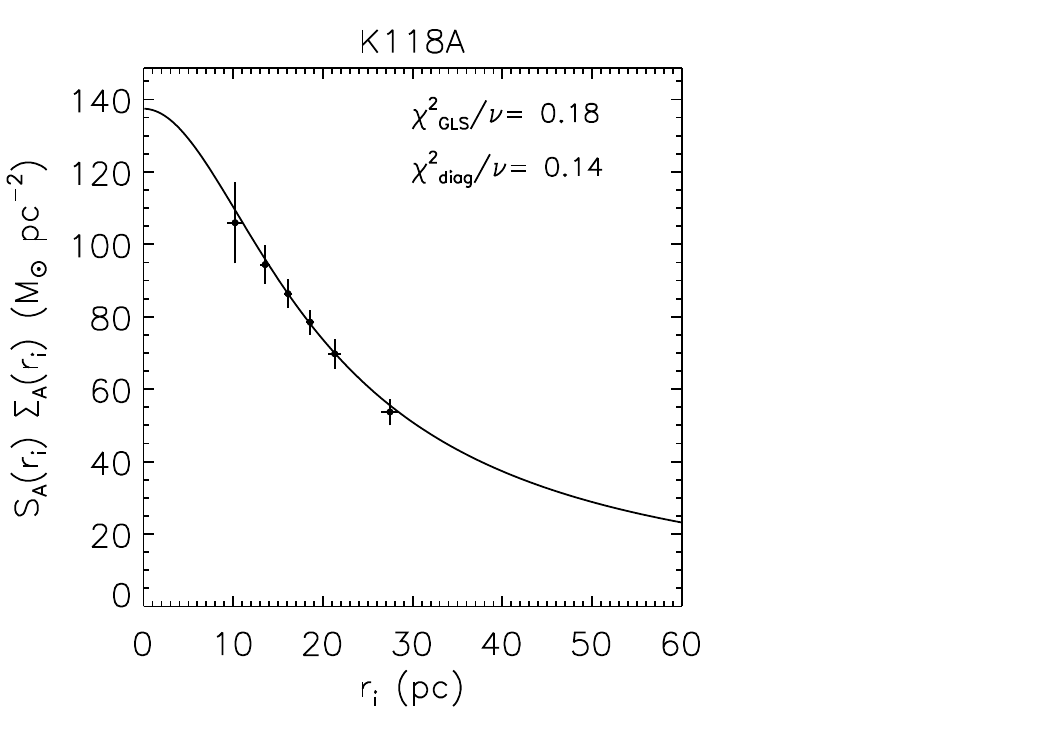} &
\includegraphics[width=2.5in,trim={0 0.0in 0.0in 0.0in},clip]{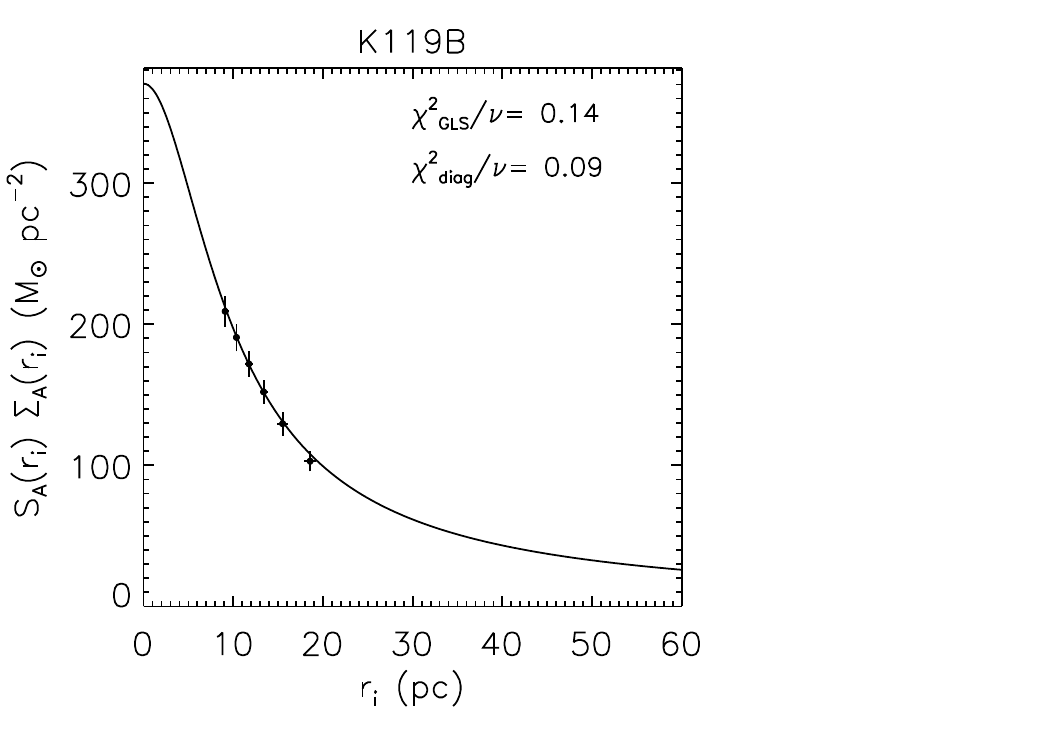} \\
\includegraphics[width=2.5in,trim={0 0.0in 0.0in 0.0in},clip]{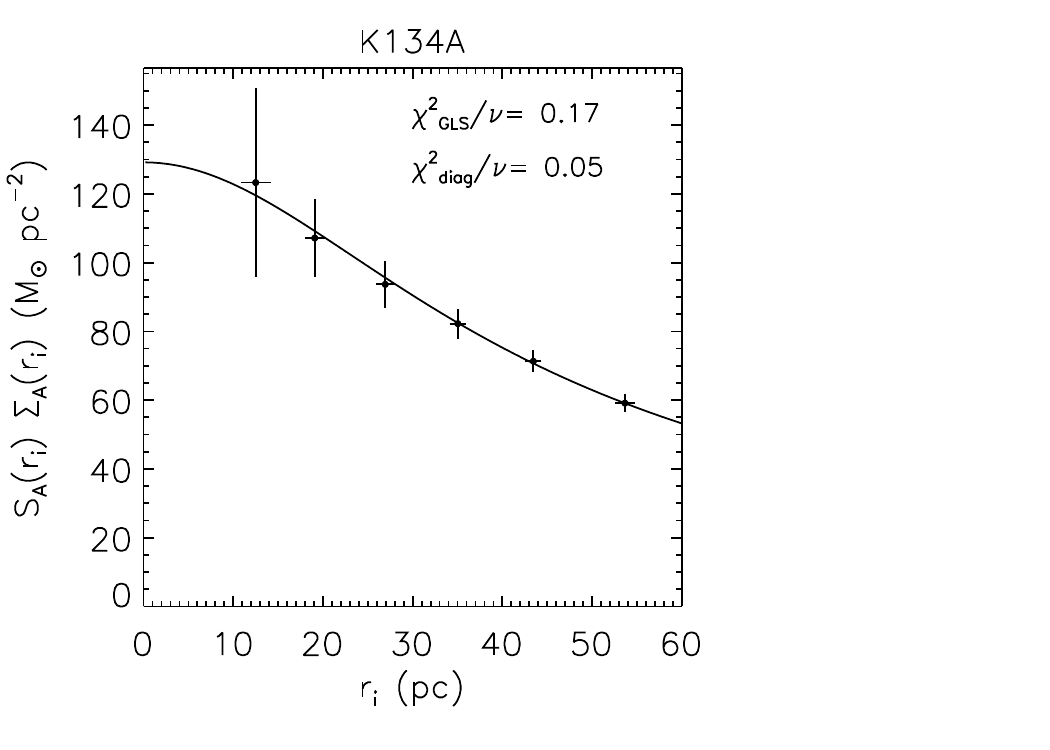} &
\includegraphics[width=2.5in,trim={0 0.0in 0.0in 0.0in},clip]{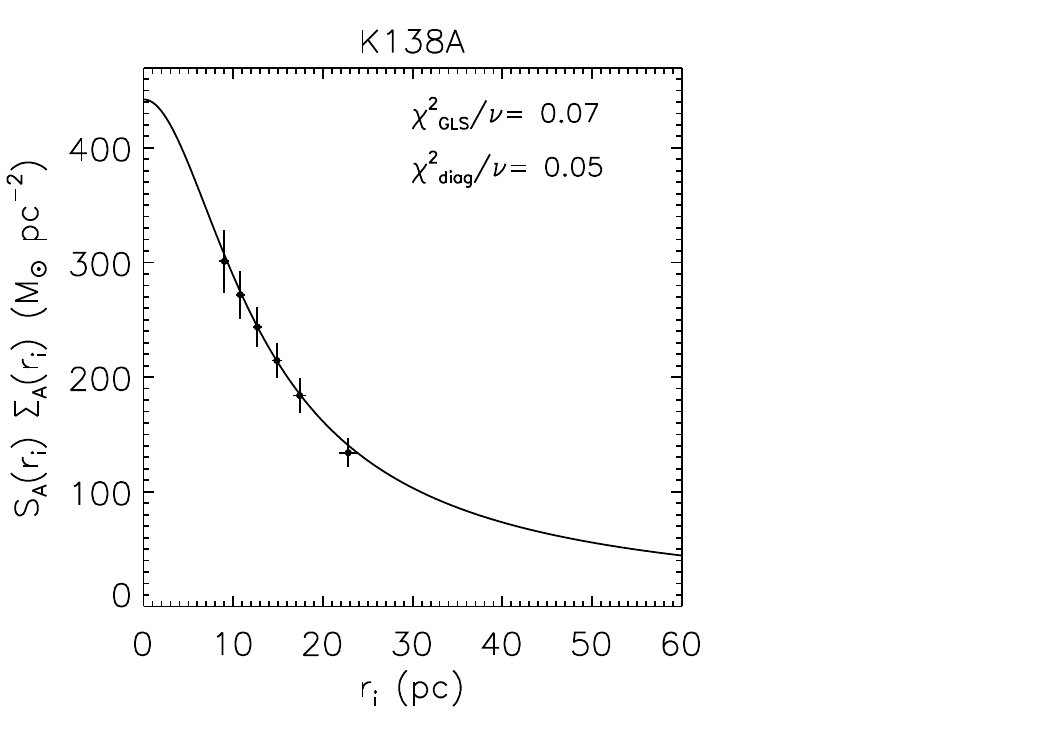} &
\includegraphics[width=2.5in,trim={0 0.0in 0.0in 0.0in},clip]{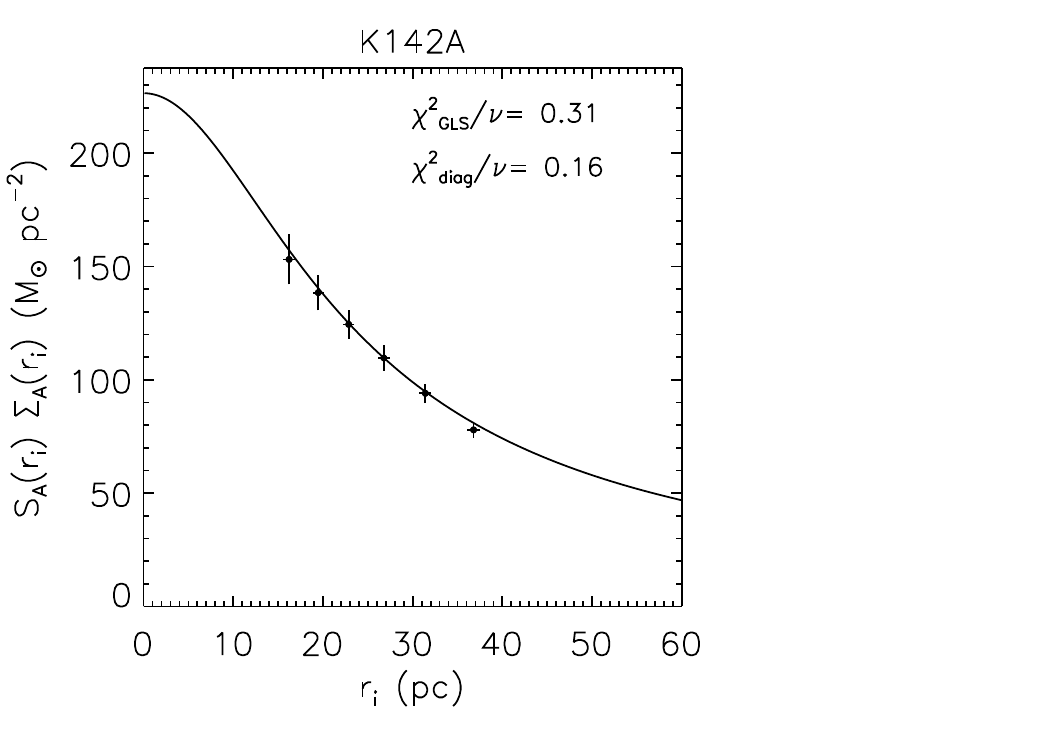} \\
\end{tabular}
\caption{
Comparison of the observed area-based surface density profiles, $\Sigma_A(r_i)$ (points) with the 
surface density profiles $S_A(r_i)$ (line)  of the Lane-Emden equation converted to physical units.
The plot shows only every other data point for clarity.
Error bars show the $1\sigma$ statistical uncertainties. 
The reduced generalized least squares $\chi^2_\mathrm{GLS}/\nu$ value and the weighted
least squares $\chi^2_{\rm diag}$ from the diagonal elements of the covariance matrix are shown for each cloud.
}
\label{individual_profiles}
\end{figure*}

\begin{figure*}
\begin{tabular} {lcr}
\includegraphics[width=2.5in,trim={0 0.0in 0.0in 0.0in},clip]{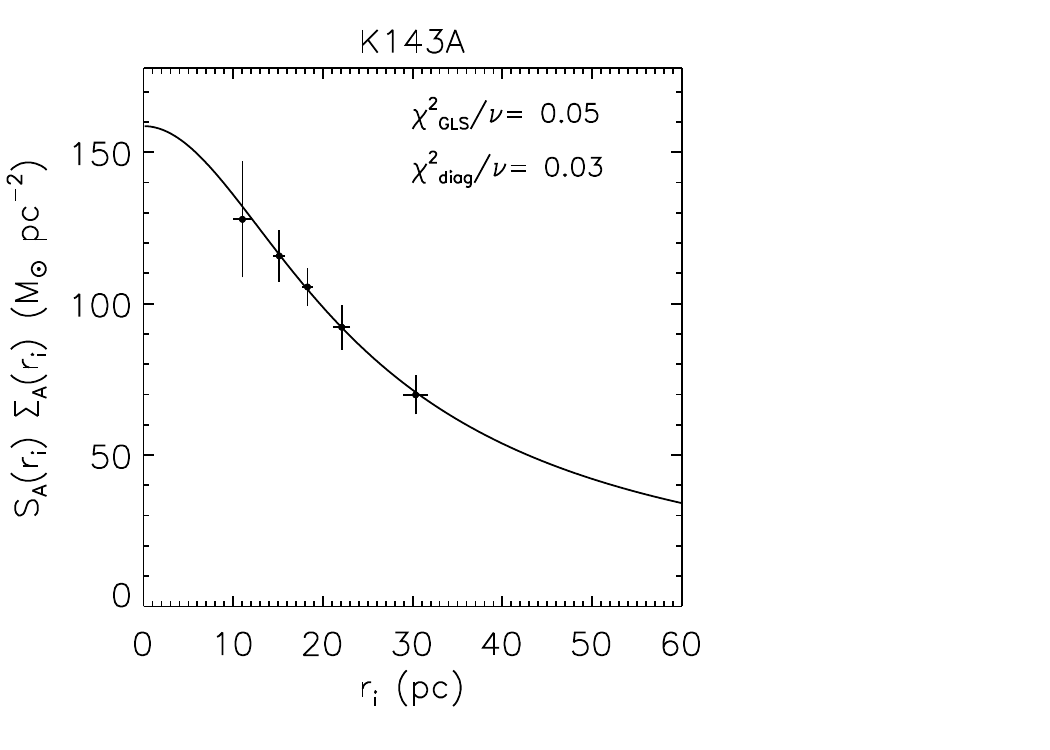} &
\includegraphics[width=2.5in,trim={0 0.0in 0.0in 0.0in},clip]{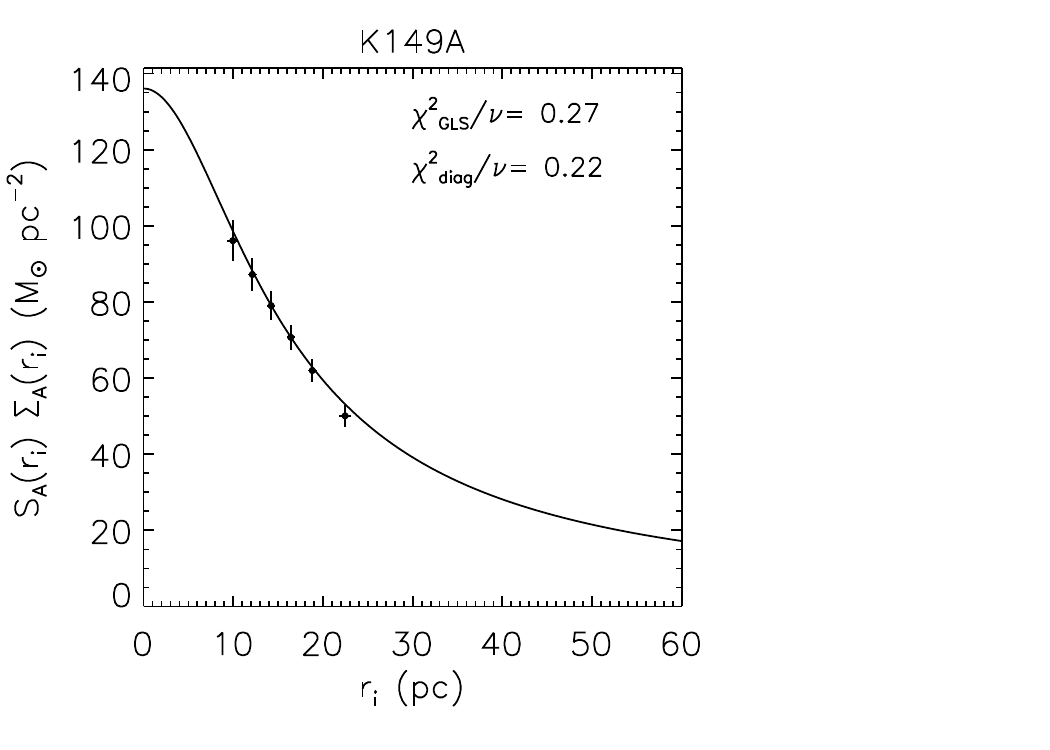} &
\includegraphics[width=2.5in,trim={0 0.0in 0.0in 0.0in},clip]{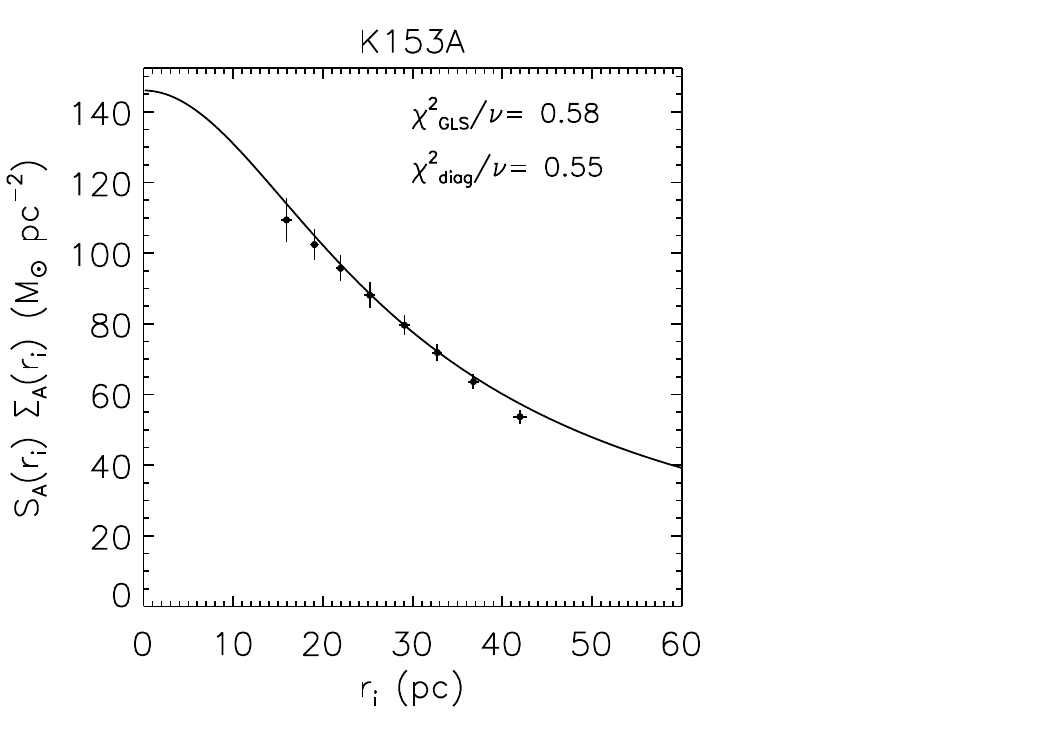} \\
\includegraphics[width=2.5in,trim={0 0.0in 0.0in 0.0in},clip]{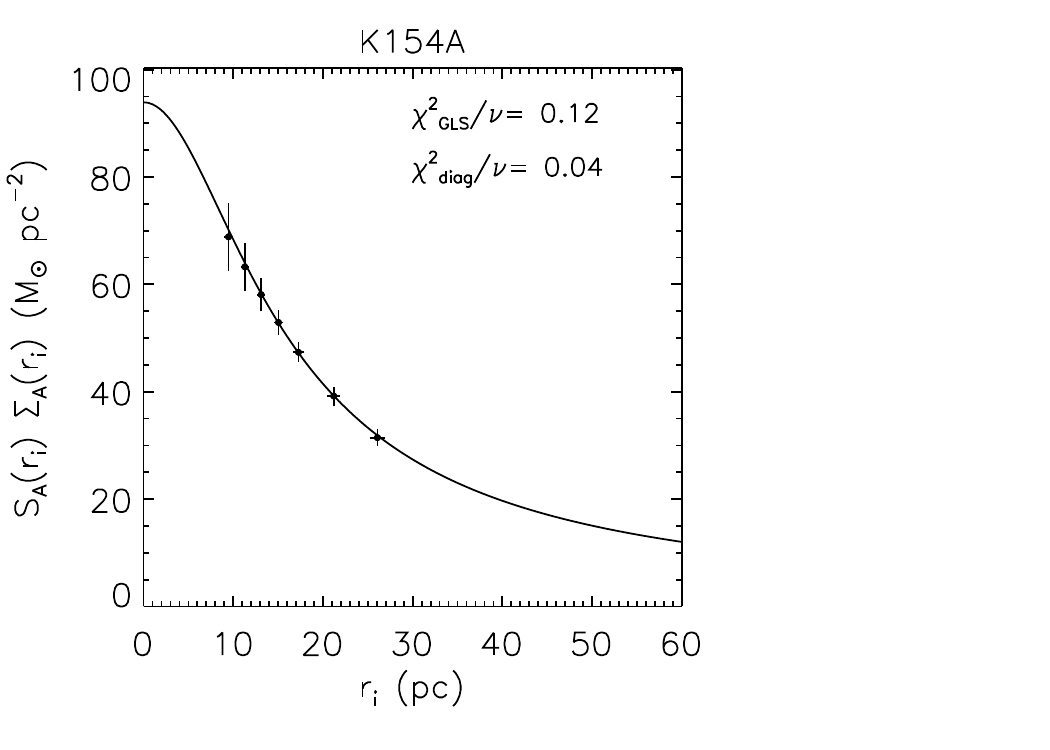} &
\includegraphics[width=2.5in,trim={0 0.0in 0.0in 0.0in},clip]{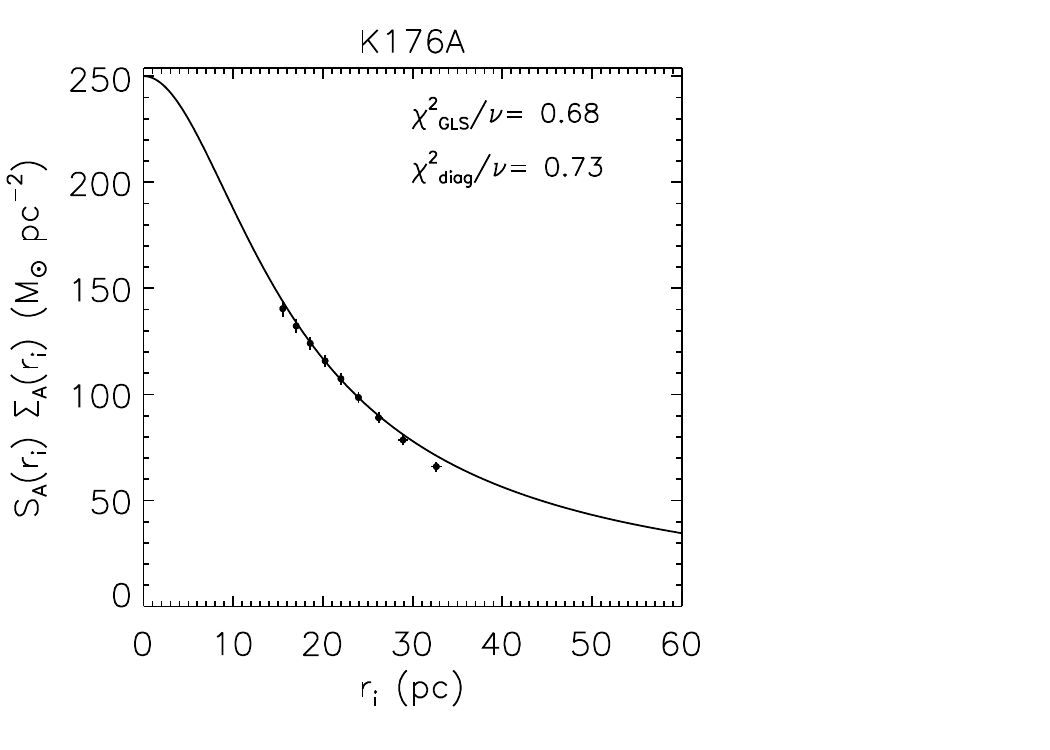} &
\includegraphics[width=2.5in,trim={0 0.0in 0.0in 0.0in},clip]{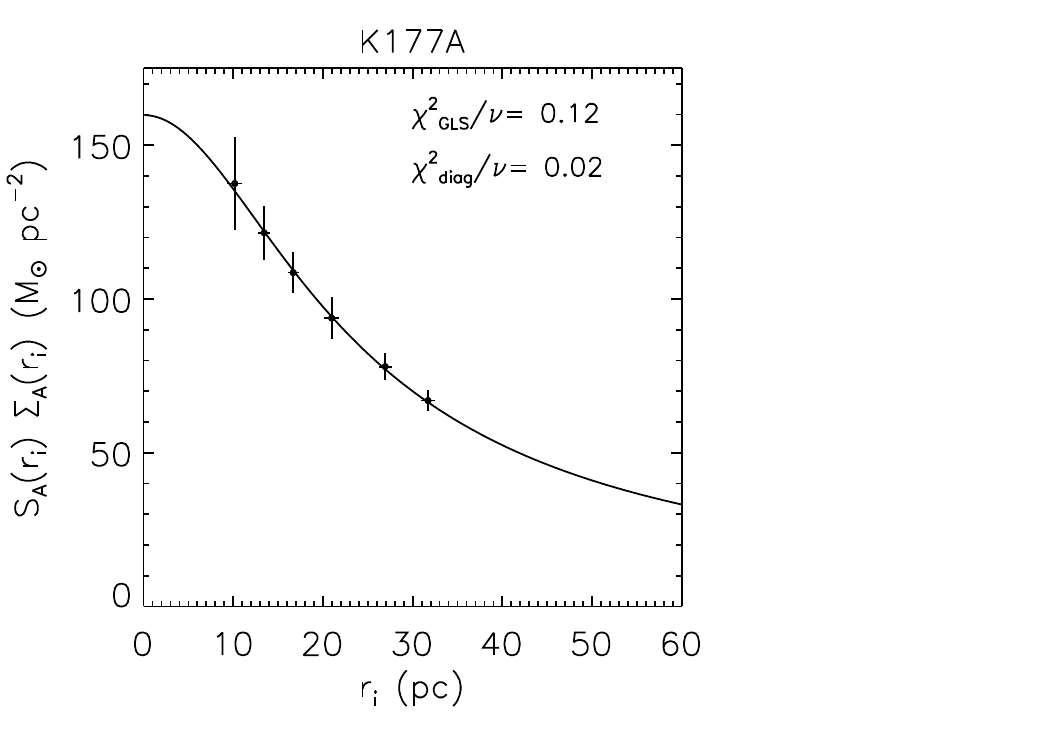} \\
\includegraphics[width=2.5in,trim={0 0.0in 0.0in 0.0in},clip]{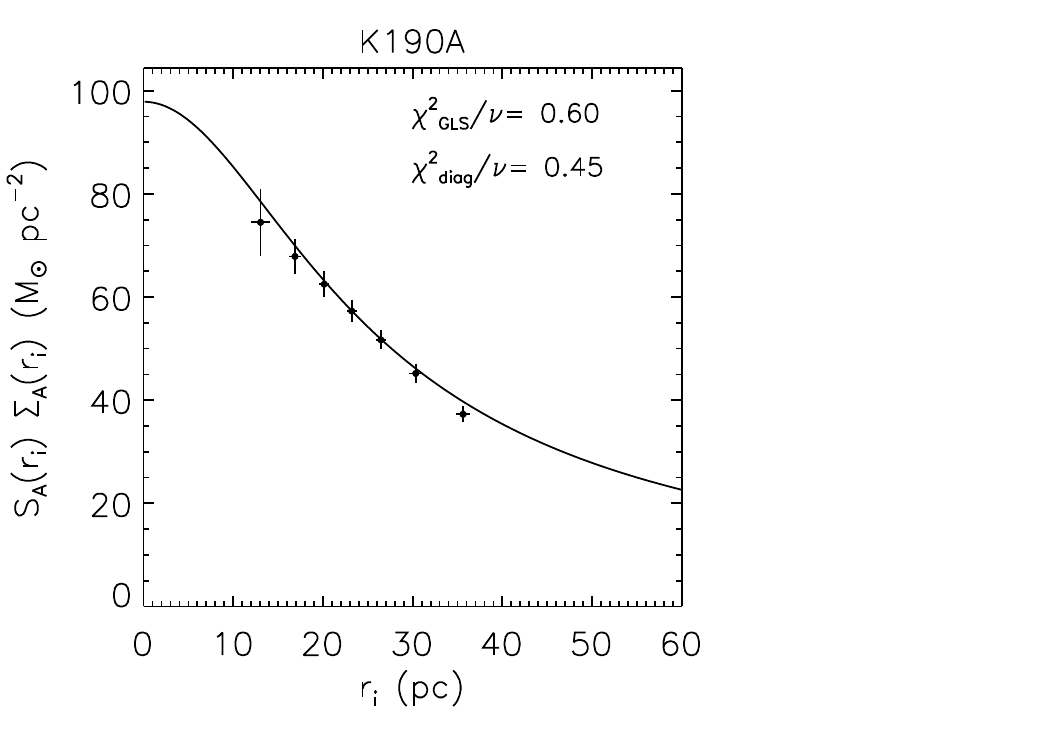} &
\includegraphics[width=2.5in,trim={0 0.0in 0.0in 0.0in},clip]{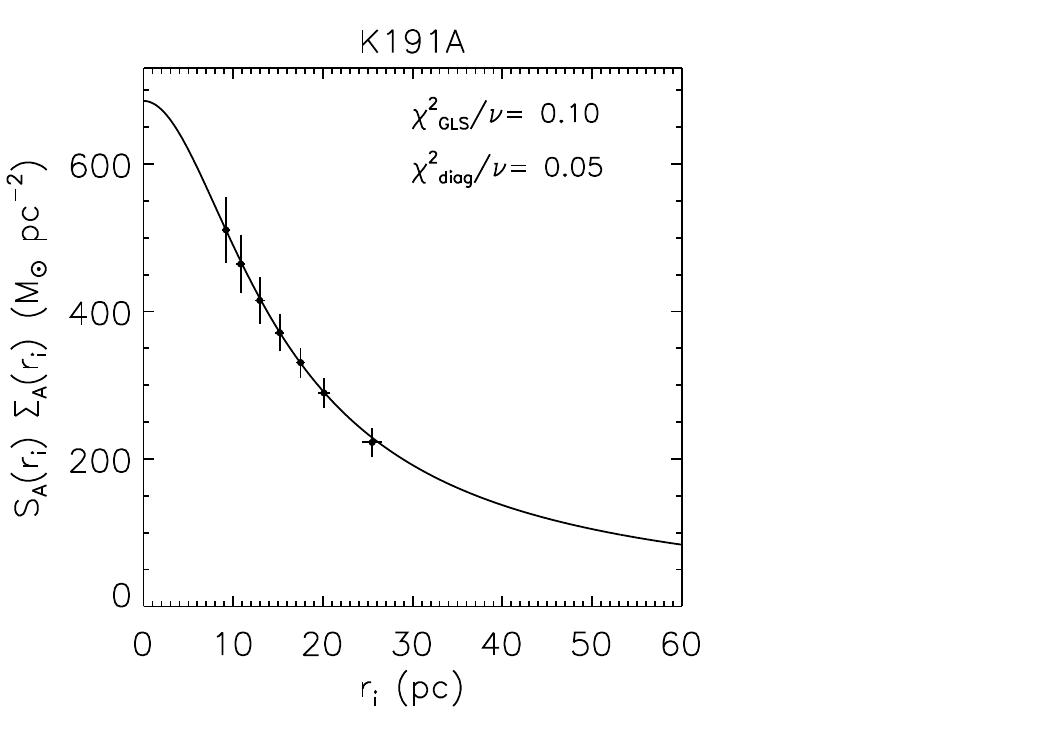} &
\includegraphics[width=2.5in,trim={0 0.0in 0.0in 0.0in},clip]{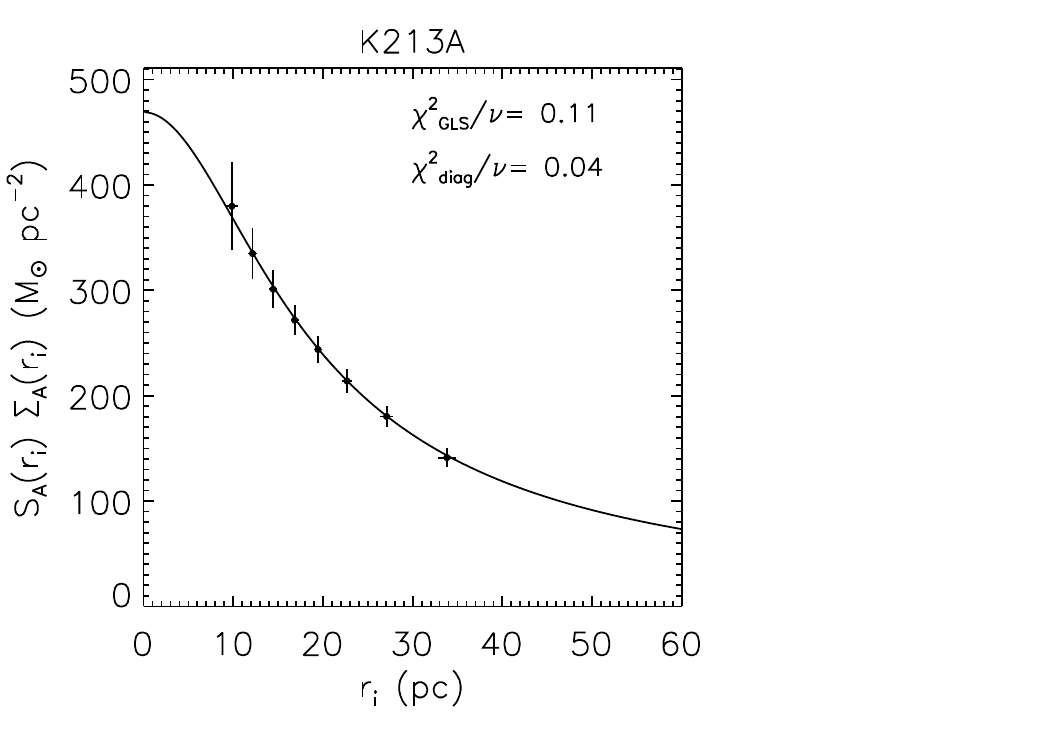} \\
\includegraphics[width=2.5in,trim={0 0.0in 0.0in 0.0in},clip]{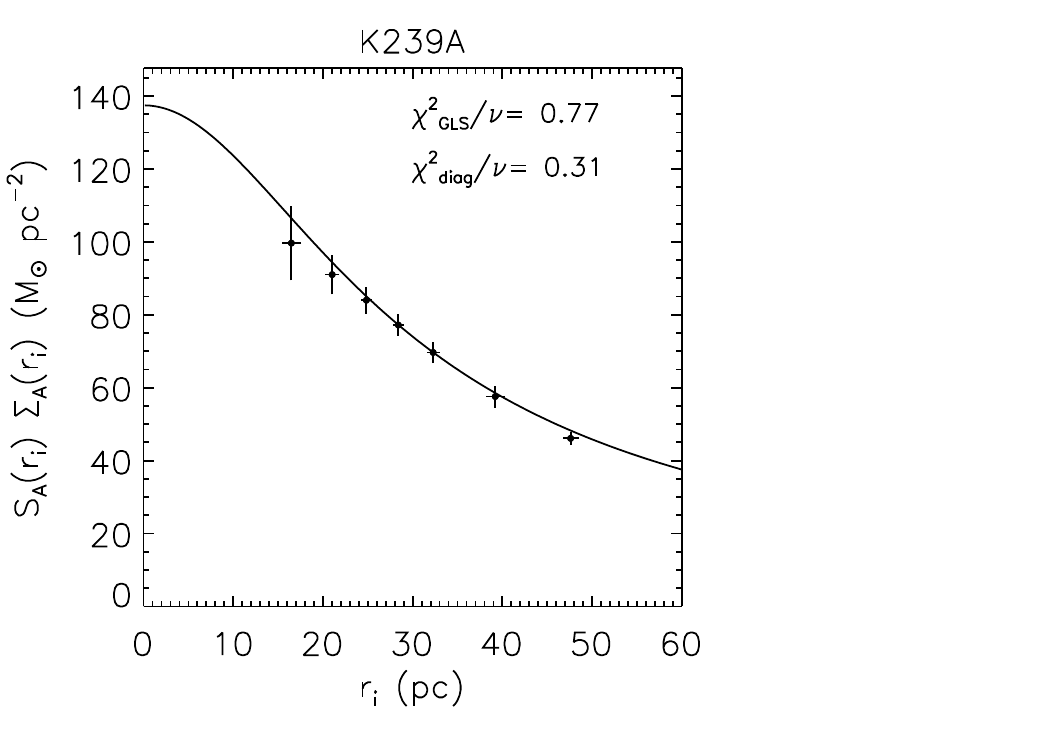} &
\includegraphics[width=2.5in,trim={0 0.0in 0.0in 0.0in},clip]{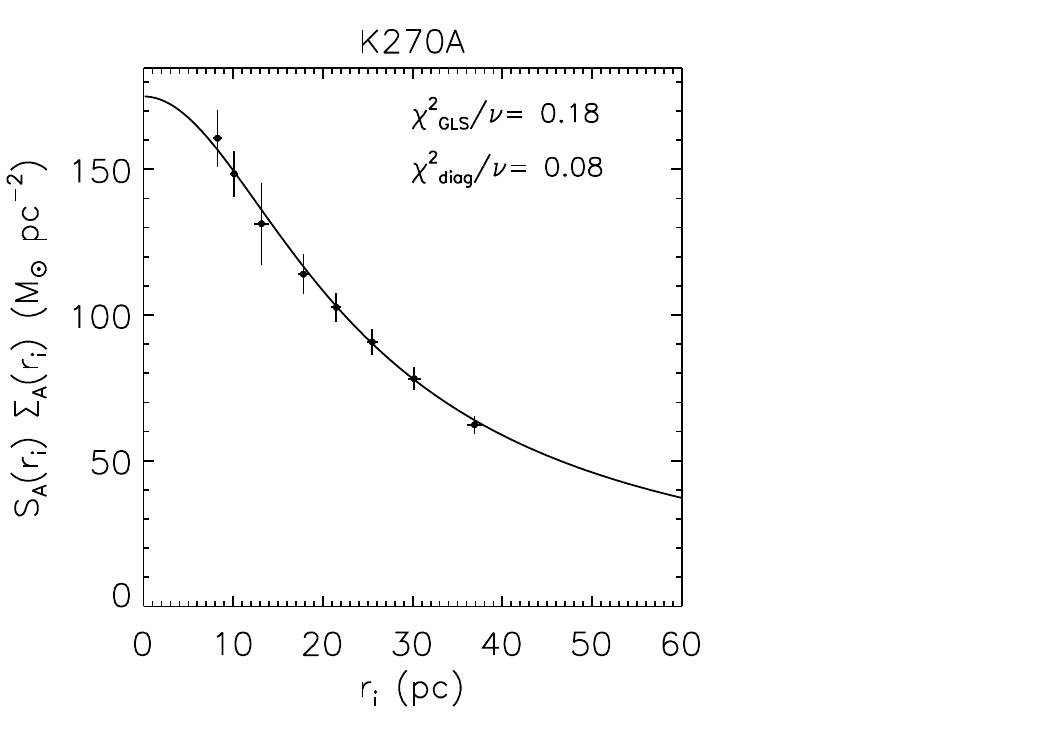} &
\includegraphics[width=2.5in,trim={0 0.0in 0.0in 0.0in},clip]{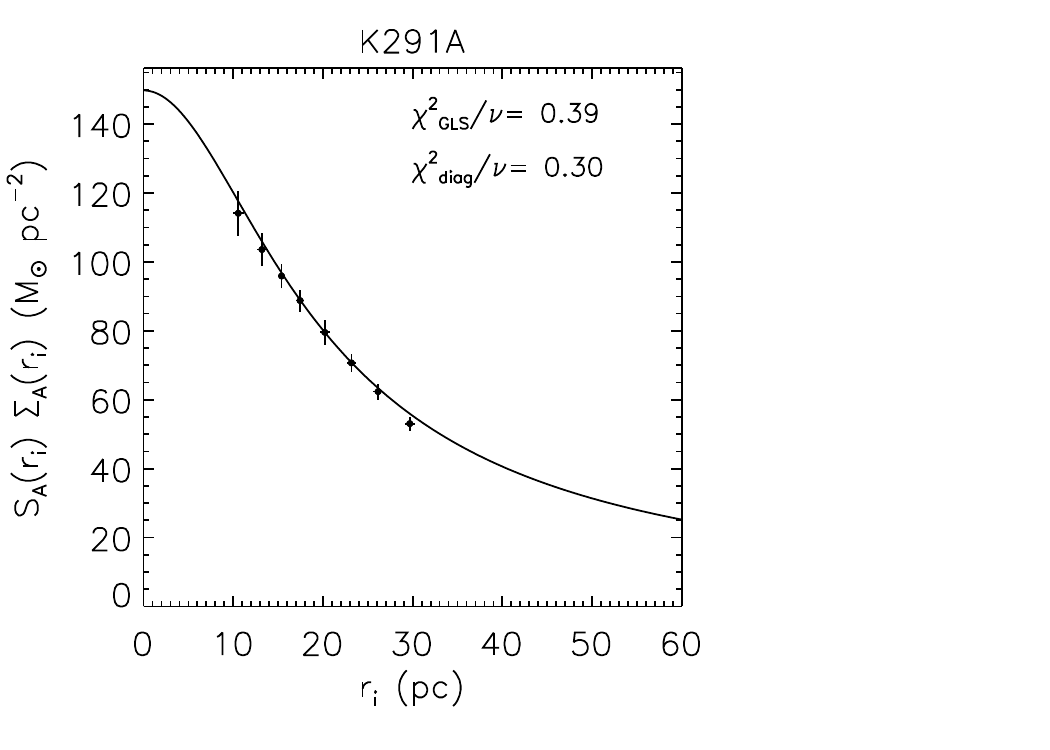} \\
\end{tabular}
\caption*{
Figure~\ref{individual_profiles} (continued)
}
\end{figure*}


 \end{document}